\documentstyle[twoside,fleqn,aip2col]{article}
\newcommand{\AmS}{{\protect\the\textfont2
  A\kern-.1667em\lower.5ex\hbox{M}\kern-.125emS}}

\input epsf
\newcommand{\psfile}[1]{
  \setlength{\epsfxsize}{\columnwidth}
  \epsfbox{#1}\vspace{-15pt}
}

\def\journal#1&#2(#3){\begingroup \let\journal=\dummyj@urnal
    \unskip, \sl #1\unskip~\bf\ignorespaces #2\rm
    (\afterassignment\j@ur \count255=#3), \endgroup\ignorespaces }
\def\j@ur{\ifnum\count255<100 \advance\count255 by 1900 \fi
          \number\count255 }
\def\dummyj@urnal{%
    \toks@={Reference foul up: nested \journal macros}%
    \errhelp={You forgot & or ( ) after the last \journal}%
    \errmessage{\the\toks@ }}
\def\apsjournal#1&#2,#3(#4){\unskip, #1\ {\bf #2}, #3 (19#4)\unskip}

\def\mathrm{\rm}
\newcommand{\ra}{\mbox{$\rightarrow$}}

\newcommand{\ee}{\mbox{${e^{+}e^-}$}}
\newcommand{\swsq}{\mbox{${\sin\theta_W}$}}
\newcommand{\mz}{\mbox{${M_Z}$}}
\newcommand{\mw}{\mbox{${M_W}$}}
\newcommand{\gz}{\mbox{${\Gamma_Z}$}}
\newcommand{\pt}{\mbox{${p_{\tau}}$}}
\newcommand {\bea} {\begin{eqnarray}}
\newcommand {\eea} {\end{eqnarray}}
\newcommand {\beq} {\begin{equation}}
\newcommand {\eeq} {\end{equation}}

\title{
\begin{flushright}
SLAC-PUB-7654 \\
September 26, 1997\\
\end{flushright}
\center {\bf Experimental Studies of Electroweak Physics}}
\author{\center {\bf Erez Etzion} 
\address {\it Department of Physics, University of Wisconsin,
Madison, WI 53706, USA.\\
Mailing address: Stanford Linear Accelerator Center, 
Stanford University, 
Stanford, CA 94309.\\
Erez@SLAC.Stanford.EDU }
\thanks{Invited talk at ``Fundamental Particles and Interactions'', 
Vanderbilt University, Nashville, Tennessee, May 1997}
\thanks{Work supported by Department of Energy contracts
DE-AC02-76ER00881 (Wisconsin) and DE-AC03-76SF00515 (SLAC).}
}

\begin{document}

\begin{abstract}
Some experimental new Electroweak physics results measured at 
the LEP/SLD and the TEVATRON are discussed.
The excellent accuracy achieved by the experiments still yield no
significant evidence for deviation from the Standard Model predictions,
or signal to physics beyond the Standard Model.
The Higgs particle still has not been discovered and a  low bound is
given to its mass.  

\end{abstract}
\maketitle

\section{INTRODUCTION}
The Electroweak (EW) theory of the Standard Model (SM)~\cite{SM} 
went through numerous tests of its predictions. The large number of free 
parameters suggest  that even if it is the correct theory, it is still 
not the final one, and there is the one missing part, the undiscovered 
Higgs  particle~\cite{Higgs}.

From the gauge structure of the theory~\cite{Taylor} and the fact that 
the W and the Z gauge bosons as well as the fermions are  massive can 
be explained by a symmetry  breaking procedure ~\cite{Higgs}. 
The easiest way this is achieved is by a doublet 
of scalar fields, which after the absorption of additional degrees of 
freedom  results in neutral physical particle, the Higgs.
This scheme can imply simple relation between the masses of the W 
and the Z gauge 
bosons and the strength of the weak couplings (which can be expressed in 
terms of the weak mixing angle) as following:
\beq
\swsq =  1 - \frac{\mz}{\mw}
\eeq
This work is a compilation of the conference 16 EW-session talks,
describing  measurements of these three parameters as well as other 
EW measurements done recently by the LEP, SLD and the TEVATRON experiments.
Some other EW results were discussed at the Heavy Flavour physics 
session~\cite{Manly}.

The next section discusses the LEP beam energy, luminosity and 
cross-section measurements. Section~\ref{sec:asym} discusses
Weak Neutral Current measurements through forward backward 
asymmetries and tau polarization measurements at LEP
and polarized asymmetries at SLD.
Charged weak couplings of the tau  measurements at LEP and SLD are
described in the following section.
Section~\ref{sec:bosons} discusses W mass measurements at LEP-2,
Trilinear Gauge Coupling and W properties measurements at the TEVATRON.
Section~\ref{sec:searches} discusses some of the searches for new 
particles at LEP-2, and the EW measurements status 
is summarized in section~\ref{conclusions}. 

Many people have contributed to these results, I would like to thank
them all, and apologize for any omissions. 
\section{LEP BEAM MEASUREMENTS}
\label{sec:beams}
\begin{itemize}
\item Determination of $E_{CM}$ at LEP-1 and LEP-2 - Paul Bright-Thomas.
\item Measuring the Luminosity at LEP-1 - Philip Hart. 
\item LEP-1 cross-sections - Marco Bigi.
\end{itemize}
\subsection{LEP Energy measurements}
\label{sec:energy}
The four LEP experiments (ALEPH, DELPHI, L3 and OPAL) have collected
4$\times 10^6$ events each during the years 1989-1995. The 1990-91 consist 
of data collected during energy scan with steps of 1~GeV each around the 
$Z^0$ peak. In 1993 and 1995 LEP scanned through 3 energies 
(peak $\pm$ 1760~MeV), while the 1992 and 1994 data were taken approximately
on the $Z^0$ resonance.
 
The Average energy is set   by the integral magnetic field
$E^{av} \propto \oint{B d \ell}$ where the  beam  path length is fixed by 
the RF frequency $\oint{B d\ell} \propto f_{RF}$. 
The storage ring transverse spin polarization (Sokolov and Ternov 
effect~\cite{sokolov}) provides the best tool for calibration of the beam 
energy using resonant depolarization of the $e^-$ beam. The intrinsic 
precision of this  method is to better than 200~KeV, however the main 
problem is that it can not be applied during normal data taking conditions,
but only in special time allocated for that. In 1995 most of the fills were
calibrated at the end of the storage ring fill, but for consistency check 
a few were also read at the beginning of the fill.

In order to obtain high precision measurement using the whole scan data, 
one would like to have continuous measurements of the $E_{CM}$ at the 
interaction points (IP).
However, one only have occasional measurements of the average energy, and 
continual measurements of the magnetic field in a few dipoles.
Therefore an extrapolation algorithm was developed to calculate the energy
while taking into account the time behavior of the magnets, current, field,
temperature as well as the geometrical properties of the LEP ring. 
It is somewhat surprising how tiny ground motions are amplified by the 
strong LEP focusing causing an energy variation of a few MeV. 
It is impressive to see how well these effects resulting from 
terrestrial tides, or even heavy rainfall and the level of water in the
lake of Geneva are nicely tracked by orbit measurements, and with other
known energy variation sources, such as temperature, day night effect etc.
are included in the energy calibration model.

Unexplained energy jumps during the 1993 run, triggered the installation of 
NMR probes inside the LEP magnets to track this 5~MeV energy variation.
The source of these unexplained jumps was identified to be the leakage
current from the .. TGV the Geneva-Bellegarde railway. 

Special conditioned at the 1995 run caused by the train-bunch mode 
operation, and the development of superconducting cavities have required
special corrections and have introduced extra small systematic 
uncertainties.

Correction due to the  beam energy spread ($55 \pm 1.5$~MeV) 
determined partly by the bunch length measured in the experiments are
 included in the Z width calculation.

The energy uncertainties are specific to the year, energy point, IP and
so on. Table~\ref{tab:lepenergy} illustrates typical energy uncertainties used 
during the 1995 run.

\begin{table}[thb]
\label{tab:lepenergy}
\begin{center}
\caption{Typical LEP energy uncertainty 
values for $E^{IP}_{CM}$ during 1995.}
\begin{tabular}{l c}
\hline \hline
Source                     & $\sigma(E_{CM}$) [MeV] \\
\hline
Depolarization             & 0.2 \\
$\ee$ difference           & 0.3 \\
Calibration statistics     & 1.0 \\
Dipole rise model          & 1.0 \\
Temperature model          & 1.0 \\
Horizontal correctores     & 0.3 \\
Tide                       & 0.4 \\
Orbit drifts               &  0.5 \\
RF corrections             & 1.0 \\
Dispersion                 & 0.7 \\
\hline
Total                      & 2.0 \\
\hline \hline
\end{tabular}
\end{center}
\end{table}

The energy uncertainties for the 1993-95 runs are
detailed in ref.~\cite{wilkinson}, it is translated
to uncertainties of 1.6~MeV on the determined
mass and width of the Z boson.
 
In June 1996 LEP-2 began to take data (10~pb$^{-1}$) at the WW threshold,
$E_{CM}=161$~GeV. Five months  later LEP energy increased to 172~GeV, 
allowing reconstruction of the W mass from its decay  products, and the
 plan is to keep ramping to 184, 196~GeV and to accumulate 500~pb$^{-1}$ 
per experiment.
The lower precision required is still very challenging, 
and techniques similar to the LEP-1 calibration are employed for LEP-2
operation mode.
The main caviate is that the
 depolarization technique is still limited to 50~GeV,
and therefore calibration is extrapolated from reading at 50~GeV to 80~GeV.
The extrapolation is the  dominant source of systematic error contributing
24/27 MeV (29/30) MeV in the 161~GeV (172~GeV) run.

\subsection{LEP-1 Cross-sections}
\label{sec:crosssec}
The results presented here are based on the 40~pb$^{-1}$ collected in 1995
combined with those recorded in the previous years. 
Hadron selection is based on high multiplicity of tracks/calorimeter
clusters, high and balanced energy deposition.
Typical relative uncertainties on the hadronic cross-sections are
$\frac{\Delta \sigma_{had}}{\sigma{had}} \approx 0.05 - 0.1 \%$ introduced
mainly by selection cuts, detector simulation, hadronization model and
resonant background. The systematics has significantly improved in the 
last year due to further studies of hadronic selection, and the improved 
theoretical Bhabha scattering calculations~\cite{bhlumi}.
The lepton selection is based on search for low track multiplicity and
back to  back topology. The electron-pair selection requires 2 high
energy  clusters (limited in $\cos \theta$ to avoid t-channel background).
The muon-pair selection require 2 high momentum tracks pointing to
minimum ionizing particles
(MIP) deposition in the calorimeters and signal in the muon chambers.
The $\tau$-pairs are typical 2 narrow, back to back low multiplicity jets.
The main background in each of the leptonic channels is contribution from
the other leptonic channels. 
A self consistency check is a comparison between
inclusive to single channels cross-sections.
Typical relative uncertainties on the leptonic cross sections are
0.3-1\%, 0.15-1\% and 0.3-0.6\% for $\Gamma_{ee}$, $\Gamma_{\mu\mu}$
and $\Gamma_{\tau\tau}$ respectively.
The total statistics of the four LEP collaborations are given in 
Table~\ref{tab:crosssec}.
Details of the individual analyses can be found in
 Ref.~\cite{lepcross1}-\cite{lepcross4}.

\begin{table}[htb]
\label{tab:crosssec}
\begin{center}
\caption{The LEP Statistics in units of $10^3$ events used for the analysis
         of the Z line shape and lepton Forward-Backward Asymmetries.}
\begin{tabular}{l c r r r r}
\hline \hline
& year &ALEPH &DELPHI &L3 &OPAL \\ 
\hline
$q\bar{q}$ & 90-91    & 451 & 357 & 416 & 454 \\
           & 92       & 680 & 697 & 678 & 733 \\
           & 93 prel. & 640 & 677 & 646 & 646 \\
           & 94 prel. & 1654 & 1241 & 1307 & 1524 \\
           & 95 prel. & 739 & 584 & 311 & 344 \\ \hline
           & Total    & 4164 & 3556 & 3358 & 3701 \\ \hline \hline
leptons    & 90-91    & 55 & 36 & 40 & 58 \\
           & 92       & 82 & 70 & 58 & 88 \\
           & 93 prel. & 78 & 74 & 64 &82 \\
           & 94 prel. & 190 & 135 & 127 & 184 \\
           & 95 prel. & 80 & 67 & 28 & 42 \\ \hline
           & Total    & 485 & 382 & 317 & 454 \\
           \hline \hline
\end{tabular}
\end{center}
\end{table} 

The measurement of the hadronic cross-section around the Z peak 
allows determination of the Z mass and width, and the peak hadronic
cross-section. Measurement of the leptonic cross-sections allows 
determination of $R_\ell=\Gamma_{had}/\Gamma_{\ell}$ for each or all leptons
combined. However it has become customary to include the lepton asymmetry 
measurements in a global fit machinery which will be discussed in 
section~\ref{conclusions}.

\subsection{LEP-1 Luminosity measurements}
\label{sec:lum}
An important aspect of the line shape measurement is related to $Z^0$
decays into invisible channels. Measurement of 
the ratio of Z decay width into invisible particles and the leptonic width,
$\frac{\Gamma_{inv}}{\Gamma_{\ell^+\ell^-}}=
\frac{\Gamma_Z-3\Gamma_{\ell^+\ell^-}-\Gamma_{had}}{\Gamma_{\ell^+\ell^-}}$
provide an interesting window to new physics due to its 
relative insensitivity to top and Higgs masses or QCD corrections.
With the large statistics accumulated at LEP this measurement was limited
by the experiments measured luminosity uncertainty (0.5\%). The method
chosen by the LEP experiments to measure the luminosity is to use the
well calculated (0.1\%) low-angle t-channel Bhabha scattering process.
The differential cross-section that is proportional to $\theta^{-3}$ 
gives a large statistics, enable to use small detector that is separated
from the other part  of the detector. The detectors can take advantage of the 
LEP relative small beam spot, well understood energy behavior 
(section~\ref{sec:energy}), and the  low backgrounds.
DELPHI  uses a sampling calorimeter with tungsten mask to define
a volume, L3 combine their crystal forward detector with 3 planes of 
silicon to define their acceptance where OPAL and ALEPH are use
silicon detector for position discrimination interleaved with tungsten
radiator for energy determination.
The preliminary LEP luminosity uncertainties are  given in 
Table~3.
%\ref{tab:lum}

\begin{table}[htb]
\label{tab:lum}
\caption{Preliminary LEP luminosity uncertainties and the 
current theoretical (Monte Carlo) limitation.}
\begin{center}
\begin{tabular}{l c c c c}
\hline \hline
 & \multicolumn{4}{c}{$\Delta L / L \times 10^{-4}$} \\ \hline
 & \multicolumn{3}{c}{Experimental} &  Theoretical \\ \hline
 & 1993 & 1994 & 1995 & \\ \hline 
ALEPH & 8.7 & 7.3 & 9.7 &  16 \\ 
DELPHI & 24 & 9 & 9 & 11-16 \\ 
L3 & 10 & 7.8 & 12.8 & 11 \\ 
OPAL & 4.6 & 4.6 & 4.6 & 11 \\ \hline \hline 
\end{tabular} 
\end{center} 
\end{table} 
 
Combining the LEP line shape results one gets  
$\Gamma_{\ell^+\ell^-}=83.89 \pm 0.11$~MeV, $\Gamma_{had}=1743.5 \pm 2.4$~MeV
and $\Gamma_{inv}=499.8 \pm 1.9$~MeV. 
With that one can derive  
$\Gamma_{inv}/\Gamma_{\ell^+\ell^-}=5.957 \pm 0.022$ compared to the
SM prediction ($n/3 \times 5.973 \pm 0.003$), or the number of light neutrinos
$N_{\nu}=2.992 \pm 0.011(exp.) \pm 0.005 (M_t,M_H)$.
Alternatively, one can assume 3 neutrino species to extract 95\% C.L. 
upper limit on additional invisible decays of the $Z^0$,  
$\Delta\Gamma_{inv} < 2.9$~MeV.  

\section{ASYMMETRY MEASUREMENTS AND NEUTRAL CURRENT COUPLINGS} 
\label{sec:asym} 
\begin{itemize}
\item Measurements of the leptonic FB asymmetries 
at LEP-1 - Wenwen Lu
\item Fermion-pair cross-sections and Asymmetries at LEP-2 -
Thorsten Siederburg
\item Spin analysis of $\ee \ra \tau^+\tau^-$ at LEP - Reinhold Volkert
\item Left-right Asymmetry measurement at SLD - Henry Band
\item Leptonic couplings Asymmetries with polarized Z -Michael Smy
\item LEP hadronic charge asymmetry - Pascal Perrodo
\end{itemize}
Around the $Z^0$ peak the fermion-pair are produced mainly through
the $Z^0$ channel, where the $\gamma$ exchange contribution is very 
small. Asymmetry measurements, Forward-Backward (FB) and polarized
asymmetries are sensitive to the right handed $Zff$ couplings 
complementary  to the partial widths measurements which are more
sensitive to the left handed couplings.
\bea
\label{eq:af}
\Gamma_f & \propto g^2_{Lf}+g^2_{Rf} & \propto v^2_f +a^2_f \\
A_f       =\frac{\sigma^L_f-\sigma^R_f}{\sigma^L_f+\sigma^R_f}
          &= \frac{g^2_{Lf}-g^2_{Rf}}{g^2_{Lf}+g^2_{Rf}}
           &= \frac{2v_fa_f}{v^2_f+a^2_f} \nonumber 
\eea
Here $g_{L(R)f}$ are the left (right) handed couplings, and
$v_f$ and $a_f$ are the effective vector and axial-vector $Zff$ 
couplings.

For unpolarized beams (LEP) the Forward Backward asymmetry,
\beq
A^f_{FB} = \frac{3}{4}A_eA_f,
\eeq
is sensitive to the
(initial) electron and the outgoing fermion couplings
to the $Z^0$.

For the $\tau$ lepton one can measure its polarization through the 
angular distribution of its decay products. Measuring the polarization as 
a function of the $\tau$ polar angle,
\beq
\label{eq:pt}
p_{\tau}=-\frac{A_{\tau}(1+\cos^2\theta)+2A_e\cos\theta}
                  {1+\cos^2\theta+2A_eA_{\tau}\cos\theta},
\eeq
enable to
derive both the the electron and the $\tau$ coupling to the $Z^0$
separately.

Given the longitudinal polarization of the electron beams at SLD,
one can use that knowledge to simply measure the difference
between left and right handed cross-section, 
\beq
\label{eq:alr}
A_{LR}=\frac{\sigma_L-\sigma_R}{\sigma_L+\sigma_R}=P_eA_e,
\eeq
where
$P_e$ is the polarization of the incident $e^-$ beam.
One can also measure the FB polarized asymmetry,
\beq
\label{eq:afbpol}
A^{pol(f)}_{FB}=
\frac{(\sigma_{L,F}-\sigma_{R,F})-(\sigma_{L,B}-\sigma_{R,B})}
     {(\sigma_{L,F}-\sigma_{R,F})+(\sigma_{L,B}-\sigma_{R,B})}=
\frac{3}{4}P_eA_f.
\eeq

While the Asymmetries expected from neutrinos, charged leptons, 
u-type quarks and d-type quarks are: 1, 0.15, 0.67 and 0.94 
respectively,
the sensitivity of these to the weak mixing angle, $\frac{\delta A_f}
{\delta \swsq}$ are 0, -7.9,   -3.5 and -0.6. 
For comparison all the LEP and SLD asymmetries are given in terms
of the effective mixing angle which is defined as:
\beq
\sin^2\theta^{eff}_W \equiv \frac{1}{4}(1-\frac{v_e}{a_e}),
\eeq 
where the $v_e/a_e$ is extracted from the asymmetry measurements.

Where Heavy Flavour measurements ($R_{b/c}, A_{b/c}$) test the SM
through vertex corrections, the leptonic asymmetries are sensitive
to the oblique radiative corrections. 

\subsection{Leptonic Forward Backward asymmetries at LEP-1}
\label{sec:FBLEP1}
\begin{figure}[t]
  \psfile{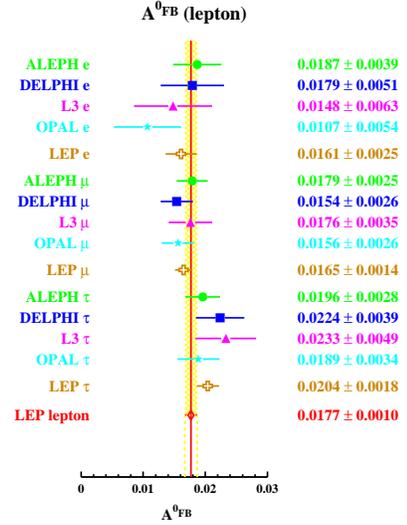}
  \caption{\label{fig:asym} Summary of the 4 LEP experiments e, 
$\mu$ and $\tau$ 
FB asymmetries.  }
\end{figure}

Two main techniques been used to measure the FB asymmetry: fitting 
the differential cross section $d\sigma/d\cos\theta \propto
1+\cos^2\theta+\frac{8}{3}A_{FB}\cos\theta$, and counting
$A_{FB}=\frac{N_F-N_B}{N_F+N_B}$.
With about $4\times400K$  lepton events at LEP, $A_{FB}$ measurements
have achieved precision of $10^{-3}$.
The LEP-1 leptons FB asymmetries are shown in Fig.~\ref{fig:asym}.
%%%%%

\subsection{Cross section and FB asymmetries at LEP-2}
\label{sec:xsecLEP2}
The fermion-pair measurements at 
LEP-2~\cite{LEP2fermion} can achieve a better, less
model dependent determination of $\mz$, and cross check lepton
universality. Compared with LEP-1 it is characterized by lower 
cross section (about two order of magnitude) 
and large radiative corrections dominated by Initial State
Radiation (ISR, so called "the return to the Z").
The differential cross-section of fermion-pair production at LEP-2 is:
\bea
&\frac{d\sigma}{d\cos\theta} = \frac{4\pi \alpha^2}{3} \{
  \frac{(\frac{\alpha(s)}{\alpha}Q_f)^2}{s}\frac{8}{3}(1+\cos^2\theta) \\  
  &+ \frac{s-\mz^2}{(s-\mz^2)^2+(\gz\mz)^2}
   [J^f_{tot}\frac{8}{3}(1+\cos^2\theta)+J^f_{FB}\cos\theta] \nonumber \\
  &+ \frac{s}{(s-\mz^2)^2+(\gz\mz)^2}
    [R^f_{tot}\frac{8}{3}(1+\cos^2\theta)+R^f_{FB}\cos\theta] \} 
\nonumber
\eea
Where
\bea
&R^f_{tot} = \frac{9}{\alpha^2\mz^2}\Gamma_e\Gamma_f \;\;;\;\;\;\;  
& J^f_{tot} = \frac{G_F\mz^2}{\sqrt{2}\pi\alpha}v_ev_f \nonumber \\
&\Gamma_f = \frac{G_F\mz^3}{6\pi\sqrt{2}}(v_f^2+a_f^2)  \\
&R^f_{FB} = \frac{3G^2_F\mz^2}{8\pi\alpha^2}v_ea_ev_fa_f \;\;;\;\;\;\;
& J^f_{FB} = \frac{3\sqrt{2}G_F}{8\alpha}a_ea_f \nonumber
\eea
and the   total cross section is:
\beq
\sigma_{tot}^f = \frac{4\pi\alpha^2}{3}\{\frac{(\alpha(s)Q_f)^2}{s}
   +\frac{J^f_{tot}(s-\mz^2)+R^f_{tot}s}{(s-\mz^2)^2+(\gz\mz)^2}
\eeq

At the $Z^0$ pole the line shape measurements were less sensitive to
the $\gamma$ propagation and $\gamma$-Z interference term,
hence the combined LEP results were
quoted with  $J^{had}_{tot}$ fixed to its SM value.
For a comparison the following are the LEP-1 quoted
values for the Z mass
\bea
\mz  = 91186.3 \pm 1.9 \;\;\;\;& J^{had}_{tot}\;\; a\;\; free\;\; parameter,
\nonumber \\
\mz  = 91189.3 \pm 6.2  \;\;\;\;& J^{had}_{tot}\;\; a\;\; fixed\;\;SM \;\;
Value \nonumber
\eea
However LEP-2 energy allows one to improve the $\mz$ determination
while $J^{had}_{tot}$ kept unconstrained $\mz$~(LEP-2)~$=91187.6\pm3.3$.

\begin{figure}[t]
  \psfile{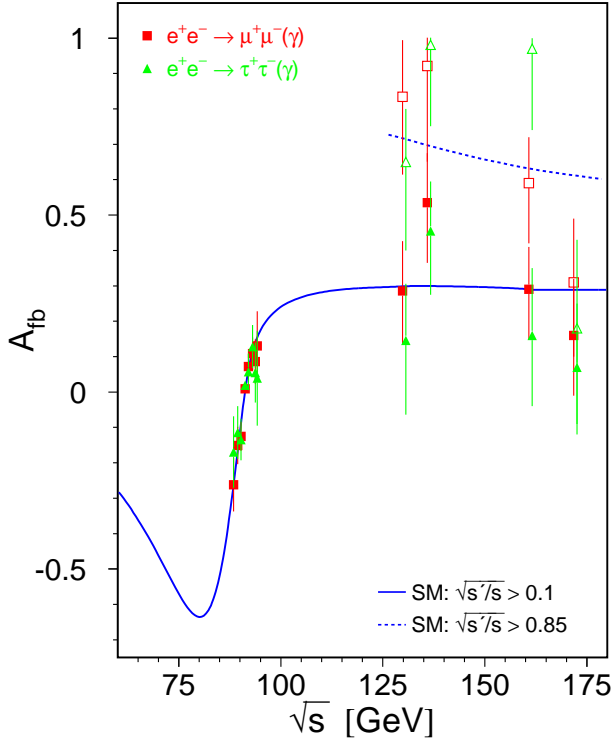}
  \caption{\label{fig:asym2}  L3 $\mu$ and $\tau$ pair FB asymmetries at
LEP-1 and LEP-2 energies, $\sqrt{s}$ is the $E_{CM}$, where 
$s^{\prime}$ is corrected for the "return for the Z" radiation 
$s^{\prime}=s-2E_{\gamma ISR}\sqrt{s}$  }
\end{figure}

Fig.~\ref{fig:asym2} shows L3 muon and tau pairs FB asymmetry results
at LEP-1 and LEP-2 energy compared with the SM predictions.
The measured cross-sections and leptonic FB asymmetries agree well
with the SM predictions.

\subsection{Tau Polarization}
\label{sec:pt}
All LEP experiments extract separate tau polarization ($\pt$) 
by reconstructing the decay of the tau to the five different channels:
$e\nu_{\tau}\bar{\nu_e}$, $\mu\nu_{\tau}\bar{\nu_{\mu}}$, $\pi\nu_{\tau}$,
$\rho\nu_{\tau}$ and $a_1\nu_{\tau}$
~\cite{pt1}-\cite{pt4}. While the first three are one
dimensional fit, in order to obtain maximum information from 
the $\rho$ and the $a_1$ channels one has to investigate the
distribution of the charged and neutral $\pi$s. With that
the $a_1$ channel becomes more sensitive than the leptonic channels
where the $\rho$ and the $\pi$ are the most sensitive channels, 
contributing weights of about 40\% each in the average.
Measuring the angular dependence of the $\pt$ 
(Eq.~\ref{eq:pt}) provides nearly
independent determination of the $\tau$ and electron coupling asymmetries, 
$A_{\tau}$ and $A_e$.
Not all the LEP 90-95 has been analyzed, OPAL (final), L3 and DELPHI 
(both preliminary) have
analyzed data up to 1994 where ALEPH has published data up to 1992 only.
The LEP results are given in Table~4 and they are consistent
with lepton universality giving on average $A_{\tau}=0.1401 \pm 0.0067$,
$A_e=0.1382 \pm 0.0076$ and assuming $e-\tau$ universality
$A_{\ell}=0.1393 \pm 0.0050$.
\begin{table}[htb]
\label{tab:pt}
\begin{center}
\caption{LEP results for $A_{\tau}$ and $A_e$. The $\chi^2/d.o.f$ 
for the average is 1.1/3 and 1.8/3 respectively.}
\begin{tabular}{l c c}
\hline \hline
Exp.          & $A_{\tau}$           & $A_e$ \\ \hline
ALEPH  & $ 0.136 \pm 0.012 \pm 0.009$  & $ 0.129 \pm 0.016 \pm 0.005 $ \\
DELPHI & $ 0.138 \pm 0.009 \pm 0.008$  & $ 0.140 \pm 0.013 \pm 0.003 $ \\
L3     & $ 0.152 \pm 0.010 \pm 0.009$  & $ 0.156 \pm 0.016 \pm 0.005 $ \\
OPAL   & $ 0.134 \pm 0.009 \pm 0.010$  & $ 0.129 \pm 0.014 \pm 0.005 $ \\
\hline \hline
\end{tabular}
\end{center}
\end{table}

For a more complete determination of the spin related structure of the
$Z^0 \ra \tau^+\tau^-$ process, ALEPH, DELPHI and L3 have provided an
independent check measuring $C_{TT}$ and $C_{TN}$ the
transverse-transverse and transverse-normal spin correlations. The
measurement is based on full angular distribution of the decay product
(excluding the $\nu$, hence it is not necessary to reconstruct the
$\tau$ direction) where different decay  combinations have different
sensitivities. The LEP average values are $C_{TT}=0.98\pm 0.11$ and $C_{TN}=0.02\pm
0.13$ in good agreement with the SM predication of $\approx 0.99$ and
$\approx 0.025 - 0.012$

\subsection{Polarized asymmetries measurements}

SLD has a new preliminary measurement of $A_{LR}$ (Eq. ~\ref{eq:alr}) based on the data
collected in 1996.
The event sample, mostly consists of hadronic $Z^0$ decays,
has 28,713 and 22,662 left- and right-handed electrons respectively.
The resulting measured asymmetry is thus
$A_m = (N_L - N_R) / (N_L + N_R) = 0.1178 \pm 0.0044$(stat).
To obtain the left-right cross-section asymmetry at the SLC
center-of-mass energy of 91.26 GeV, a very small correction
$\delta = (0.240 \pm 0.055)\%$(syst)
is applied which takes into account residual contamination in the
event sample and slight beam asymmetries.
As a result,
\bea
  A_{LR}(91.26~\mbox{GeV}) &=&
       \frac{A_m}{\langle P_e \rangle} \left(1 + \delta\right)
       = 0.1541 \\ \nonumber
        & &\pm 0.0057 \mbox{(stat)} \pm 0.0016 \mbox{(syst)}
\eea
where the systematic uncertainty is dominated by the systematic
understanding of the beam polarization.
Finally, this result is corrected for initial and final state radiation
as well as for scaling the result to the $Z^0$ pole energy:
\begin{eqnarray}
  A^0_{LR} & = & 0.1570 \pm 0.0057 \mbox{(stat)} \pm 0.0017 \mbox{(syst)}
 \\ \nonumber
  \sin^2\theta_W^{eff} & = & 0.23025 \pm 0.00073 \mbox{(stat)}
       \pm 0.00021 \mbox{(syst)}.
\end{eqnarray}
The 1996 measurement combined with the previous (published) measurements
 yield:
\begin{eqnarray}
  A^0_{LR} & = & 0.1550 \pm 0.0034 \label{Equ_ALR}\\ \nonumber
  \sin^2\theta_W^{eff} & = & 0.23051 \pm 0.00043.
\end{eqnarray}
Which is  the single most precise determination of
weak mixing angle.

SLD has presented a direct measurement of the $Z^0$-lepton coupling
asymmetry parameters based on a sample of 12K leptonic $Z^0$ decays
collected  in 1993-95~\cite{afbpol}. The couplings are extracted from the measurement
of the double asymmetry formed by taking the difference in number of
forward and backward events for left and right beam polarization data 
samples (Eq. ~\ref{eq:afbpol}) for each lepton species.
This measurement has a statistical advantage of 
$(P_e/A_e)^2\approx 0.25$ on the LEP FB asymmetry measurements. It
is  independent of the SLD $A_{LR}$ using $Z^0$ decays to hadrons, and
it is the only measurement which determines $A_{\mu}$ not
coupled to $A_e$.
The results are: $A_e=0.152\pm 0.012(stat)\pm 0.001(sys)$, 
$A_{\mu}=0.0102\pm 0.034\pm 0.002$, and $A_{\tau}=0.195\pm 0.034\pm
0.003$ or assuming universality $A_{\ell}=0.151\pm0.011$. 

The SLD  preliminary weak mixing angle value combining $A_{LR}$, $Q_{LR}$ and
leptons asymmetries measurements is:
\begin{equation}
  \sin^2\theta_W^{eff} = 0.23055 \pm 0.00041.
\end{equation}
which is more than 3$\sigma$ below the LEP average.

\subsection{The Hadronic Charge Asymmetry ${\langle Q_{FB} \rangle}$}
\label{QFB}
The LEP experiments~\cite{QFB1}-~\cite{QFB4}  have provided 
measurements of the average charge flow 
in the inclusive samples of hadronic decays which is  related to FB of the
individual quarks asymmetry as following:
\beq
<Q_{FB}>=\sum_{quark\;\;flavour}{\delta_fA_{FB}^f
          \frac{\Gamma_f}{\Gamma_{had}}}.
\eeq
The charge separation, $\delta_f$, is the average charge difference between
quark and antiquark in an  event. The $b$ and $c$ are extracted 
from  the data, the $\delta_b$ as a by-product of the $b$ asymmetry 
measurement (self calibration)  where the charm separation is obtained 
using the  hemisphere opposite to a fast $D^{*\pm}$. Light quark 
separations are derived from MC hadronization models which is the main 
systematic source.
The results expressed in terms of the weak mixing angle are:
\bea
0.2322 \pm 0.0008(stat) \pm 0.0007(sys.\;exp) \pm 0.0008 (sep.)
\nonumber \\
0.2311 \pm 0.0010(stat) \pm 0.0010(sys.\;exp) \pm 0.0010 (sep.)
\nonumber \\
0.2336 \pm 0.0013(stat) \pm 0.0014(sys.\;exp) \;\; (New !)
\nonumber \\
0.2321 \pm 0.0017(stat) \pm 0.0027(sys.\;exp) \pm 0.0009 (sep.)
\nonumber 
\eea
for ALEPH, DELPHI, L3 and OPAL respectively.

%%%% taulorentz etc.

\section{TAU LORENTZ STRUCTURE}
\begin{itemize}
\item Measurements At LEP - Joachim Sommer
\item Measurements At SLD - Erez Etzion
\end{itemize}
Measurements of the Charged Current structure in the leptonic and 
semi-leptonic decays of the tau  searching for deviation from  
V-A which was measured
to a good accuracy at the muon sector.
The measurements of the 4 Michel parameters ($\rho$, $\eta$, 
$\xi$ and $\xi\delta$)
in the leptonic decays, and $\xi_{had}\sim h_{\nu}$, 
the helicity of the tau neutrino in the hadronic decays are 
extracted from reconstructed kinematic parameters of the decay particles. 
The $\rho$ and $\eta$ parameters are measured from the leptonic decays 
energy spectrum, were the correlation between the two taus is utilized in 
the LEP experiments (ALEPH, OPAL, l3) as well as CLEO and ARGUS
for the determination of $\xi$, $\xi\delta$ and $h_{\nu}$.
SLD exploits the high polarization of the 
incident electron beam to extract these quantities directly from a
measurement of the tau decay spectra.
The results are still not as precise as in the muon sector, but
are also consistent with the $V-A$ SM prediction.
Table~5
%\ref{Tab:lorentz} 
summarizes the various  measurements.

\begin{table*}[htbp]
% space before first and after last column: 1.pc
% space between columns: 2.0pc (twice the above)
  \setlength{\tabcolsep}{1.pc}
\label{Tab:lorentz}
\caption{Summary of present present experimental measurements and SM predictions for the $\tau$ Charged Current parameters.} 
\begin{tabular*}{\textwidth}{l c c c c c }
\hline \hline
exp.     & $\rho$          & $\eta$          & $\xi$ 
& $\xi\delta$  & $h_{\nu}$\\ \hline
SM       & $0.75$          & 0               & 1 & 0.75 & 1\\ \hline
ARGUS    & $0.738\pm0.038$ & $0.03\pm0.22$   & $0.97\pm0.14$
& $0.65\pm0.12$   & $1.017\pm0.039$\\
CLEO     & $0.747\pm0.012$ & $-0.015\pm0.08$ & $1.007\pm0.043$
& $0.745\pm0.028$ & $1.03\pm0.07$\\
ALEPH 92 & $0.751\pm0.045$ & $-0.04\pm0.19$  & $1.18\pm0.16$
& $0.88\pm0.13$   & $1.006\pm0.037$\\
ALEPH  & & & & & \\
(stat.)  & $0.749\pm0.019$ & $0.047\pm0.08$    & $1.032\pm0.077$
& $0.79\pm0.052$  & $0.996\pm0.008$\\ 
L3        & $0.794\pm0.05$  &$0.25\pm0.2$      & $0.94\pm0.22$
& $0.81\pm0.15$    & $0.97\pm0.054$\\ 
SLD       & $0.72 \pm0.09$ & $-0.6\pm0.9$ ($\mu$)            & $1.05\pm0.35$  
& $0.88\pm0.27$    &$0.93\pm0.11$\\  
OPAL      & & & & & $1.29\pm0.28$ \\  \hline
Average & $0.748\pm0.009$   &$0.028\pm0.051$   & $1.017\pm0.035$    
& $0.761\pm0.023$ & $0.997\pm0.008$\\  
\hline \hline 
\end{tabular*} 
\end{table*} 

Due to the $\frac{M_{\ell}}{M_{\tau}}$ factor current experiments are not
sensitive to  $\eta_e$. A few experiments 
"improved" $\eta_{\mu}$ measurement using the high $\rho - \eta$ 
correlation and assumed $e - \mu$ universality.
However this is an {\it inconsistent} assumption because generally a 
non zero $\eta$ would imply non universal $\rho$.

ALEPH has  presented a new method that includes the reconstruction of 
the $\tau$ direction.
L3 published a global analysis with 50\% of their data. 
OPAL measured the $\xi_{had}$  in the
$\tau^- \ra \pi^-\pi^-\pi^+\nu_{\tau}$ as part of the measurement of the 
hadronic structure function.  
SLD published their results for 1993-95 data~\cite{sldtau}.
Most of the LEP data still haven't been analyzed, 
and SLD is still taking data, so we do expect (and there is room for)
improvements from the Z machines and  maybe more than that
from CLEO with its very large $\tau$-pairs sample.

\section{GAUGE BOSONS PROPERTIES}
\label{sec:bosons}
\begin{itemize}
\item W Mass Measurements at LEP-2 - Carla Sbarra.
\item Trilinear Gauge Couplings (D0, CDF) - Chris Klopfenstein.
\item W physics at the TEVATRON - Arie Bodek.
\end{itemize}

\subsection{W mass measurements at LEP-2}
The W discovery  and first studies of its  properties all  took  place
at  $p\bar{p}$  collisions. 
In 1996 all 4 LEP collaborations have used two complementary methods to 
determine $M_W$:
With the data recorded at 161~GeV (just above the W pair threshold) the 
measured cross-section was compared with the predicted 
one~\cite{wlep1}. The statistics is rather  low due to the low cross section
. Systematic error is dominated by the background rejection.
The LEP average  cross-section is:
$\sigma_{W^+W^-}=3.69\pm0.45$~pb and the mass derived using an average 
energy of $161.33\pm0.05$ is:
$M_W=80.40^{+0.22}_{-0.21}\pm0.03$~GeV.

At 172~GeV the cross section is about three times larger and each 
experiment has 
recorded about 100 W pairs. At this energy the W mass was determined by 
directly reconstructing the invariant mass of the  W decay  
products~\cite{wlep2}. Averaging the results treating the smallest 
systematic
error as 100\% correlated yields: 
$M_W=80.37\pm0.18\pm0.05(color)\pm0.03(LEP)$, where the second error 
is due to color reconnection and the third error is due to LEP energy 
uncertainty of 30~MeV.
Combining the two methods yield LEP average of:
$M_W=80.38\pm0.14$~GeV.
\begin{figure}[t]
  \psfile{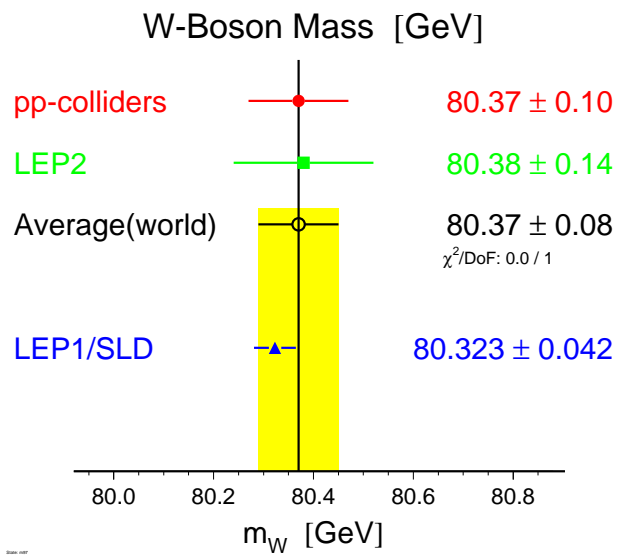}
  \caption{\label{fig:mw} Comparison of direct and indirect 
 Measurements of the W mass.
The $p\bar{p}$ average represents the results of UA2, CDF and D0,
  where the LEP results is the combined 161 and 172 GeV runs. The
  indirect is calculated using LEP-1, SLD and $\nu$-N results.}
\end{figure}

\subsection{Trilinear Gauge Couplings measurements at the TEVATRON}
\begin{figure}[t]
  \psfile{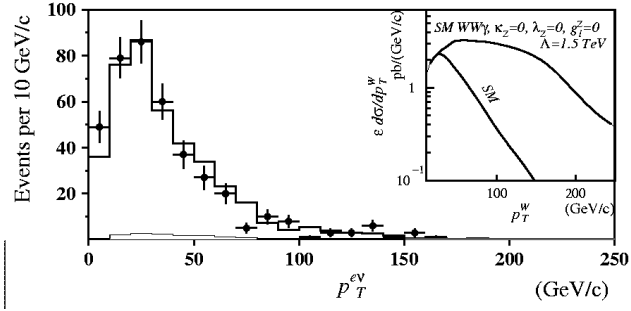}
  \caption{\label{fig:tgc} $P_T$ distribution of the $e\nu$ system
for the 1993-95 D0 data set. The points represent
the data while the solid line stands for the total
background estimate plus the SM prediction of WW
and WZ production (shown as shaded
histogram). The inset shows the predicted
$d\sigma/dp^W_T$, folded with detection
efficiencies, for SM WW$\gamma$ and WWZ couplings (lower curve), and
for SM WW$\gamma$ and anomalous WWZ couplings
(upper curve).}
\end{figure}

 The vector boson trilinear couplings predicted by the non-Abelian
gauge symmetry of the SM can be measured directly in pair production
processes such as $q\bar{q} \ra W^+W^-,\;\;W^{\pm}\gamma,\;\;Z\gamma$
or $W^{\pm}Z$.
Deviation from the SM couplings would signal new physics. 
CDF and D0~\cite{tgc} have searched for $W\gamma, WW, WZ$ and $Z\gamma$ 
diboson final states,
using several different techniques with high transverse momentum 
events of the TEVATRON, the 1.8 TeV $p\bar{p}$ collider. $WWZ$ and 
$WW\gamma$ vertices have been observed and direct limits on
 non-SM, three-boson, $WWZ$ and 
$WW\gamma$ anomalous couplings  were set.
Fig.~\ref{fig:tgc} shows no clear difference between 
the transverse momentum distribution for WW
and WZ candidate events from D0 1993-95 data to the total background estimate
plus SM expectations.
Limits were set on the
$Z\gamma$ couplings while searching for the non-SM $ZZ\gamma/Z\gamma\gamma$
interactions. CDF  has studied the $W\gamma$ interaction where SM
predicts  no radiation at $\cos \theta^*=\pm 1/3$ ($\theta^*$ is the angle 
between the incoming quark and the photon). A hint for that zero radiation
has been observed by CDF. Limits were also set on the anomalous  $WWV$ and
 $ZV\gamma$ couplings. The results are in agreement with the 
SU(2) $\times$ U(1) model of SM electroweak interactions.

\subsection{W properties studies at the TEVATRON}

The run 1a (18~$pb^{-1}$) and 1b data (110~$pb^{-1}$) 
of the TEVATRON,  was used in 
extracting the $W$ asymmetry at CDF~\cite{bodek}.
A new technique using the silicon tracking and energy 
in the electro-magnetic
calorimeter has extended the electron rapidity coverage.
The asymmetry results are used to reduce the systematic
errors on the $M_W$
measurement (from 100 to 50 MeV), and it can be even further reduce to 
the level of 20~MeV.
W width was measured in two ways: indirectly from the W and Z 
cross sections, and indirectly from the tail of the transverse mass
distribution.

Drel-Yan dilepton production at high invariant mass yield limits
on extra $Z^{\prime}$ bosons, and place strong limits on quark substructure.
New limits on quark-lepton compositness 
scales from dilepton production were recently published~\cite{qecdf}.

\section{SEARCHES FOR NEW PARTICLES}
\label{sec:searches}
\begin{figure}[t]
  \psfile{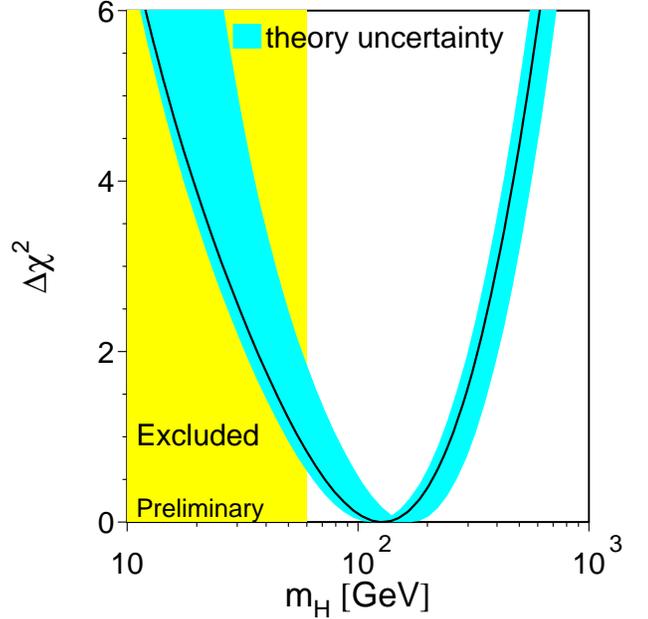}
  \caption{\label{fig:higgs}$\chi^2$ curve for the Higgs boson in the
  minimal SM fit to the winter 1997 electroweak data and the region
excluded by direct searches for the SM Higgs particle.}
\end{figure}

\begin{itemize}
\item Searches at LEP-2 - Klas Hultqvist
\end{itemize}
The increase of LEP center of mass energy has opened a new window
for new particle searches. The main missing piece in the frame 
of the electroweak puzzle is the scalar particle that breaks the 
symmetry, 
the Higgs particle, which its mass is the the only free parameter left
in the electroweak fits.
Searches at LEP-2 concentrate on looking for SM Higgs or an extension
of 2-doublet Higgs model, and "new physics"  mainly supersymmetric 
particles in the frame minimal supersymmetry model (MSSM) gauge sector 
(chargino and neutralino) and S-matter 
sector (slepton and squark),   or extensions to MSSM, and also    
for exotic particles such as excited leptons or unexpected topologies.
The Higgs reach increases with the increase of energy. Different 
topologies are used where due  to the high branching ratio of Higgs 
to $b\bar{b}$,
b-tagging plays an important role in the Higgs search.
New preliminary lower limits for the SM Higgs mass at 95\% CL are:
 L3  - 68.2~GeV/$c^2$, OPAL - 68.8~GeV/$c^2$, ALEPH
70.7~GeV/$c^2$ and last is DELPHI with a limit of 64.6~GeV/$c^2$
effected by a $h\nu\bar{\nu}$ candidate they have from the 161~GeV 
run with calculated mass of $64.6^{+5}_{-2.6}$~GeV (consistent with 
the background expectation of 1.3 events). Fig.~\ref{fig:higgs}
shows the region excluded by the SM Higgs searches at LEP compared
with the mass predicted by the the electroweak fits 
(Section~\ref{conclusions}).

The MSSM Higgs limits are model dependent and
one needs to assume for example a certain 
 sfermion mixing in setting the limits.
Typical MSSM $M_h$ are greater than 60~GeV/$c^2$.
For SUSY particles preliminary LEP-2 limits are above 84~GeV/$c^2$ for 
charginos and 25~GeV/$c^2$ for neutralinos. There are improved limits 
for sfermions as well: 63, 58 and 70~GeV/$c^2$ for the supersymmetric 
partners of the top, muon and electron. Except for ALEPH's excess of 
4 jet events at a mass of about 105~GeV/$c^2$ (discussed in~\cite{Manly})
no new particles were found and other searches have also set 
limits only on degenerate  and long live SUSY, R-parity violation, 
compositness etc.. However the expectation from the coming LEP data with
energies of 184~GeV and higher are to further improve the range for all 
searches.

 \begin{figure}[t]
  \psfile{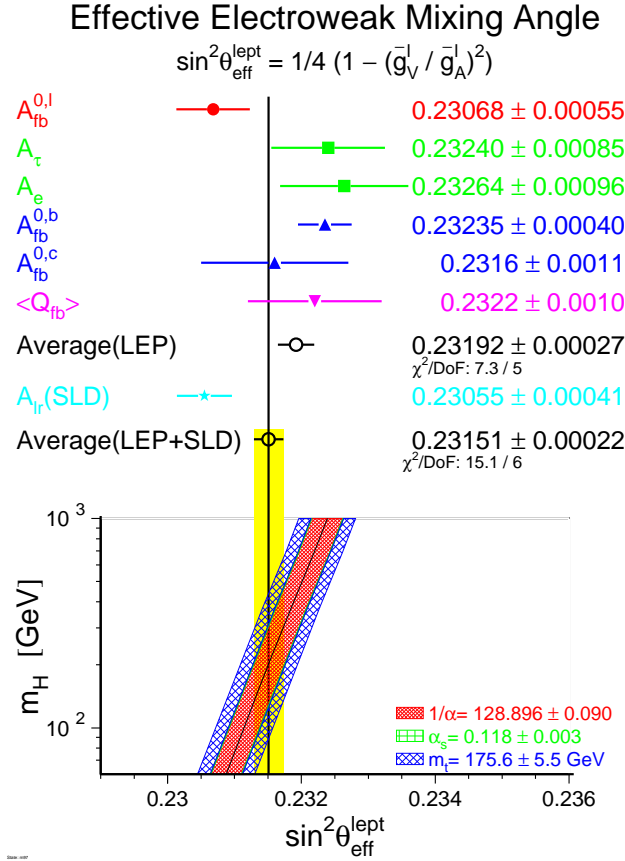}
  \caption{\label{fig:swsq} Summary of measurements of
  $\sin^2\theta^{eff}_W$
from the FB asymmetries of leptons, $\tau$ polarization, inclusive
  quarks, heavy quarks asymmetry and SLD polarization asymmetry.
Also shown is SM prediction as a function of the Higgs mass.}
\end{figure}

\section{SUMMARY} 
\label{conclusions}
The Coupling parameters $A_f$ (Eq.~\ref{eq:af}) are determined by the LEP
FB asymmetry measurements ($0.1536\pm0.0043$) and tau polarization
($0.1393\pm0.0050$) and by the polarized
measurements at SLD ($0.1547\pm0.0032$). Using $A_{\ell}$ one can
determine 
$A_{b/c}$ from LEP FB asymmetries in the heavy quarks sector. 
However one can turn it the other way around assume the hadronic 
couplings to be given by the SM and include the c/b quark asymmetry  
measurements in the determination of the effective weak mixing  angle. 
Fig.~\ref{fig:swsq} compares several determinations of the weak  
mixing angle, where the most significant disagreement is  
between LEP $A_{FB}(b)$ and $A_{LR}$ measurement at SLD.

The combination of the numerous very precise electroweak measurements 
yield stringent constraints on the SM. The data taken so far and the 
theoretical predictions agree well.  Radiative correction are well  
established as seen be the excellent agreement between the  
top mass predictions and the CDF/D0 measurements. The data show some  
sensitivity to the mass of a SM Higgs particle and the SM fits using
all the data yield
SM Higgs mass of $127^{+127}_{-72}$~GeV/$c^2$
(Fig.~\ref{fig:higgs}). One discrepancy which  
will be interesting to watch is the difference between the value of
the  weak
mixing angle as derived from $A_{LR}$ at SLD, to the one calculated at
LEP
dominated by the $b$ quark asymmetry
(Fig.~\ref{fig:swsq}). There are still many preliminary
numbers or results that are using only sub-sample of the whole LEP data
(e.g. tau polarization), and 
it would be interesting to see the final LEP numbers published along 
with the new results from LEP running in higher energies and SLD next 
year's data.  

In the $W$ physics the TEVATRON and LEP-2 are getting closer to the 
precision of the radiative corrections of the LEP-1 and SLD data,  
providing a new test of the SM (Fig.~\ref{fig:mw}). 
 
There is still no real clue for new particles and physics beyond the
SM, 
hence further analysis of the existing and future data is still  
required. 
  
\section{ACKNOWLEDGMENT} 
I  would like to thank all the speakers who contributed to this 
session, and to the LEP/SLD/Fermilab collaborations they represent  
who brought this field to that  high level of precision and interest. 
Many averages in this report were derived by the LEP/SLD  
EW group~\cite{lepewwg}. 
It is pleasure to thank the organizers of the conference for the
successful interesting meeting. 

\bibliographystyle{plain}

\end{document}